Determinants of occupational mobility within the social stratification structure in India

Vinay Reddy Venumuddala
PhD Student in Public Policy, Indian Institute of Management, Bangalore

## Introduction

Recent literature on social stratification and inequality, perceives social mobility as a phenomenon that mainly arises from the movement of individuals from their traditional social strata to new ones, within or across generations (Lambert et.al 2018, Bottero 2005, Van Leeuwen et.al 2010). Social mobility according to this line of work can be better understood only after one empirically observes the stratification patterns within a society. Importance of individual level occupation information, in observing such stratification patterns finds substantial amount of research literature in sociology. Some important works in this regard have been discussed in the works of Lambert et.al (2018) and Van Leeuwen et.al (2010). In particular, recent work by Lambert et.al (2018) illustrates various procedures by which social connections among individuals within a household can be linked to social distance between occupations, in order to identify broad occupational stratification patterns. They propose 'social resin' as a construct that helps to explain how, volume of social connections between different individuals can at an aggregated level reflect the social distance between occupations held by them. Social and cultural capital associated with the position of an occupation in the broader social space, and deep-rooted preferences for stability in individuals, are two plausible accounts which according to them drive the existence and influence of social resin. Malleability of such social resin, is what essentially accounts for social mobility, nevertheless retaining a more or less stable stratification structure over time (Lambert et.al 2018, Bottero 2005). In fact, using the information about individual occupations and social connections between them, Lambert et.al (2018), use Social Interaction Distance and Social Network Analysis based approaches to substantiate the relevance of social resin in explaining the emergent stratification patterns across countries.

In this study, we make use of empirically observed stratification patterns, in order to identify the relationship between education and social mobility of individuals; the latter is approximated by the social distance of an individual's occupation from his/her household's traditional niche. Our study draws upon a novel occupational network construction proposed in Lambert et.al (2018), with slight adjustments, to empirically identify social stratification patterns using cross sectional household surveys available in the Indian context (we use IHDS-II). As discussed before, social mobility is associated more or less with movement of individuals to occupations that are at a greater social distance from their traditional niche. Assuming that the eldest and least educated working individual is associated with a household's traditional occupation, mobility of other individuals within the household is captured by the distance of their respective occupations to the former. We measure the social distance between occupations held by individuals, using the geodesic distances between occupations from the empirically constructed occupational network[1]. The dependent variable for our study therefore is the distance of an individual's occupation relative to his/her origin (traditional occupation). We denote this variable by $d_{ij}$, for $i^{th}$ individual in $j^{th}$ household. One conspicuous limitation of our study is that variation in $d_{ij}$ doesn't account for upward or downward movements differently. While upward or downward movements can help differentiate between upward and downward mobility, scope of our study limits us to observing individual movements either closer to or distant from the corresponding traditional structural position of his/her household. Actual movements however can be observed by separately constructing occupational network from the social connections of characteristically similar groups of households that we might be interested in (for example, we can construct occupational networks separately for each social group and see where are the individuals having better educational qualifications moving to).

## Data:

We use second round of India Human Development Survey (IHDS) conducted in 2011-12. IHDS is a nationally representative large dataset covering 42152 households, with information on health, education, employment, economic status, marriage, fertility, gender relations and social capital (Desai et.al). For the purpose of network construction, we discard those occupations which are found in less than 10 households in order to avoid spurious ties.

## Occupational Network from social connections

[1] Occupations that are closer to each other, can be assumed to have similar structural positions in the broader social space. This assumption follows from the Bourdieu's conception of social space or field, which has been methodologically revived by the works of de Nooy (2003) and Lambert et.al (2017), making it possible to be observed within a social network framework. We discuss this in detail in the subsequent sections dealing with occupational network construction.

Information on individual occupations within a household, in household surveys, can help us frame a bi-modal network[2] capturing affiliations between households and occupations, which can then be projected as a unimodal network[3] of occupations (See Borgatti et.al 2018, for Bimodal to unimodal projection). This method essentially translates the social connections between individuals within a household, into social proximity between occupations. While motivation for the network construction scheme followed in Lambert et.al (2018) comes from correspondence analysis tables used within the CAMSIS social interaction distance tradition, the method we follow is of Bimodal to unimodal projection. While both the methods seem to be different, they commonly retain the intent of translating the volume of social interaction between individuals into social distance between occupations[4].

## Observed Stratification Patterns

Figure shown below indicate the empirically observed stratification pattern, depicted through demarcating community structures within an occupational network. For a better visualization we use force directed placement algorithm proposed by Fruchterman et.al (1991). Aesthetics shown is essentially as a result of algorithm trying to distribute vertices evenly in the frame, minimize edge crossings, make edge lengths uniform, reflect inherent symmetry in the graph, and conform the plot to the frame (Fruchterman et.al 1991). In order to retain the essence of proximity between the nodes, we depict the strength of connection by thickness of the edges. Clustering or community detection in this network, essentially indicates the presence of, sets of 'occupations that are at social proximity to each other', just because the social connections between individuals in different occupations within a cluster, on average, are stronger in comparison to those spanning across clusters. Communities are identified using '*cluster_louvain'* function in '*igraph*' package in R, which implements a multi-level modularity optimization[5] algorithm for finding community structures in large networks (Blondel et.al).

[2] Where the incidence matrix is constructed between households and occupations, with each cell representing whether or not any of the household members are associated with a particular occupation or not. We consider appropriate household weights while performing Bimodal to unimodal projection, details are provided in appendix.

[3] Where connections between occupations reflect the social distance. Unlike network construction scheme adopted in Lambert et.al (2017) where only marriage relationship is accounted for while observing social connection between individuals, we observe social connections between all the working individuals within a household.

[4] For methodological similarity between correspondence analysis, and use of network analysis (particularly theory related to projecting an affiliation matrix to a one-mode network) in the context of social stratification, particularly within Bourdieu's social field theory, refer, De Nooy W, 2003.

[5] Essentially, modularity optimization finds optimal community structures such that ties within any community are stronger relative to ties between communities. This however, doesn't imply that there won't be any stronger ties between nodes spanning different clusters. Since criteria for optimization involves only average strength of connections within and across communities, tie strengths between nodes are essentially treated in an aggregate manner and not considered in isolation.

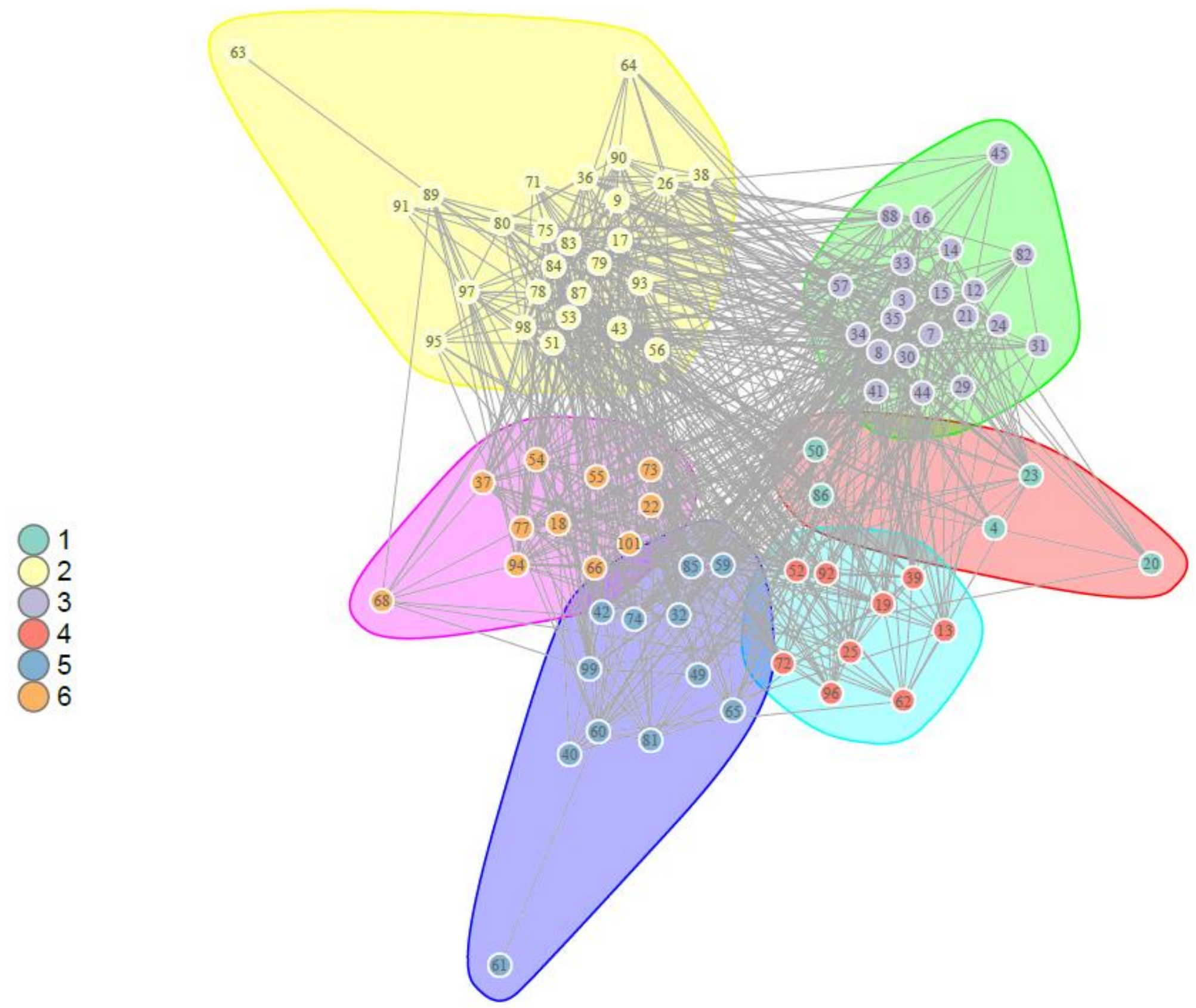

*Figure 1: Clustering pattern observed in occupational network (Note: Occupation codes are based on NCO-1968 2-digit classification)*

### Relationship between Education $(E_{ij})$ and $d_{ij}$

Effect of education on $d_{ij}$ may be confounded by the following factors out of many.

1. Fixed accounts corresponding to origin occupation which is the traditional niche within a household $(Ot_j)$, have to be accounted for. The movement of household members socially farther from origin occupation may vary depending on the origin occupation itself, sometimes irrespective of the education levels of their members. For example, it is observed from the data that social distance from agriculture labour ("63") is farther from almost all the other occupations. However, among all others we find that it is closer to Potters and related workers ("89") and construction workers ("95")[6]. Therefore, controlling for origin occupations is essential as it may explain significant portion of the observed social distance among individuals.
2. Level of education of the corresponding member associated with origin occupation $(Oe_j)$. Higher the education associated with origin occupation, it is plausible to assume that higher may be the education of its members. For observing the marginal effect of education on $d_{ij}$ it is essential therefore to control for this factor.
3. Geographical context in which the households reside $(R_j)$. Areas with better education facilities on average may cause the education of individuals on average to be higher, and therefore make it difficult to identify the ceteris paribus effect of education.
4. It can happen that female members in the households have less education and therefore less or more $d_{ij}$. Controlling for the gender of household members can therefore help us come closer to identifying the marginal effect of education on $d_{ij}$.
5. And most importantly, the relationship between education and extent of mobility is different for different social groups. Hence it is essential to control for social group fixed effects $(S_j)$.

$$d_{ij} = \alpha + \beta_0 E_{ij} + \beta_2 Oe_j + Female_{ij} + Ot_j + R_j + S_j + \epsilon_{ij}$$

[6] In the network diagram the connection between "63" and "95" is not depicted. To avoid too many lines within the network plot, we only use those edges that are above the 3rd quartile in the overall edge weight distribution.

**Results**

| | *Dependent variable:* | | | | |
|---|---|---|---|---|---|
| | Social distance from traditional occupations (dij) | | | | |
| | (1) | (2) | (3) | (4) | (5) |
| Education | 1.886*** | 1.995*** | 3.286*** | 3.354*** | 3.123*** |
| | (0.031) | (0.032) | (0.038) | (0.038) | (0.043) |
| FC | | 2.935*** | 0.081 | -0.016 | -0.155 |
| | | (0.849) | (0.823) | (0.823) | (0.818) |
| OBC | | 6.450*** | 2.131*** | 1.685** | 1.389* |
| | | (0.813) | (0.790) | (0.791) | (0.786) |
| SC | | 8.089*** | 2.592*** | 1.899** | 1.434* |
| | | (0.830) | (0.808) | (0.811) | (0.806) |
| ST | | 9.563*** | 3.475*** | 2.185** | 1.621* |
| | | (0.891) | (0.869) | (0.878) | (0.872) |
| MUS | | 9.691*** | 3.833*** | 3.454*** | 2.898*** |
| | | (0.897) | (0.874) | (0.874) | (0.872) |
| OTH | | 1.041 | -1.052 | -0.876 | -0.957 |
| | | (1.263) | (1.223) | (1.222) | (1.214) |
| Assets | | | | -0.322*** | -0.164*** |
| | | | | (0.031) | (0.032) |
| Urban | | | | | -2.926*** |
| | | | | | (0.398) |
| Female | | | | | 3.552*** |
| | | | | | (0.294) |
| Age | | | | | -0.228*** |
| | | | | | (0.011) |
| Oe | | | -2.820*** | -2.728*** | -2.599*** |
| | | | (0.049) | (0.050) | (0.050) |
| Observations | 49,376 | 49,373 | 49,373 | 49,356 | 49,356 |
| $R^2$ | 0.091 | 0.097 | 0.154 | 0.156 | 0.167 |
| Adjusted $R^2$ | 0.089 | 0.095 | 0.152 | 0.154 | 0.165 |
| Residual Std. Error | 29.885 (df = 49259) | 29.796 (df = 49250) | 28.833 (df = 49249) | 28.805 (df = 49231) | 28.617 (df = 49228) |
| *Note:* | | | | | *p<0.1; **p<0.05; ***p<0.01 |

*Figure 2: Regression results ($d_{ij}$ on Education).*

*Note: State, and $O_t$ (origin occupation) fixed effects are controlled for in the regression.*
*$O_e$ indicates the education of the household member associated with $O_t$.*

Regression results indicate that Education significantly explains variation in $d_{ij}$ and is positively related. Significant coefficient for origin education indicates that a greater marginal effect of education on mobility may be observed at lower prior education levels within the household. Higher asset levels of a household are associated with less $d_{ij}$, which indicates that individuals tend to move less far from their traditional occupations when they possess more assets. Urban, higher assets, and high level of initial education are on average associated with less movement from origin occupations, which indicates the existence of a stable stratification structure at higher levels of economic and cultural capital, constituting such households. Females are on average associated with occupations that are at a greater social distance in comparison to men, indicating the presence of a predominant gender stratification within occupations.

## Relationship with income

In the previous regression, we considered the effect of education on the movement of individuals to occupations socially distant from their traditional niche. However, we couldn't comment on the direction of such movement, whether it is for better prospects or worse. Here we try to find out if the social distance is associated positively or negatively with an approximation of economic distance between occupations. We regress MPCE of individuals on occupations as factor variable, and assign a proxy for economic capital, which equals the predicted MPCE corresponding to an occupation code. Similar to distance from traditional occupation within a household, we also construct a variable $ED_{ij}$ which indicates the economic distance of individual occupations from the corresponding origin occupation. Following regression is used to inquire into the possible factors affecting the relationship between $d_{ij}$ and $ED_{ij}$.

$$ED_{ij} = \delta + \gamma_1 d_{ij} + \gamma_2 E_{ij} + Female_{ij} + S_j + R_j + Ot_j + Oe_j + \nu_{ij}$$

All the correlates of $d_{ij}$ also are used as controls in this regression, so as to avoid inconsistencies in estimation.

**Results**

| | *Dependent variable:* | | |
|---|---|---|---|
| | Economic distance from traditional occupations (EDij) | | |
| | (1) | (2) | (3) |
| dij | 24.801$^{***}$ | 13.022$^{***}$ | -32.928$^{***}$ |
| | (0.802) | (0.826) | (1.339) |
| Education (Eij) | | 387.964$^{***}$ | 204.452$^{***}$ |
| | | (8.374) | (9.257) |
| FC | -516.385$^{***}$ | -478.715$^{***}$ | -444.640$^{***}$ |
| | (152.983) | (149.757) | (147.010) |
| OBC | -875.996$^{***}$ | -762.997$^{***}$ | -747.178$^{***}$ |
| | (147.087) | (144.003) | (141.360) |
| SC | -1,099.772$^{***}$ | -941.452$^{***}$ | -903.480$^{***}$ |
| | (150.770) | (147.628) | (144.921) |
| ST | -624.299$^{***}$ | -462.827$^{***}$ | -430.329$^{***}$ |
| | (163.195) | (159.789) | (156.858) |
| MUS | -1,020.755$^{***}$ | -572.128$^{***}$ | -503.034$^{***}$ |
| | (162.809) | (159.667) | (156.744) |
| OTH | -640.170$^{***}$ | -582.089$^{***}$ | -585.709$^{***}$ |
| | (227.205) | (222.414) | (218.331) |
| Assets | 194.965$^{***}$ | 138.507$^{***}$ | 129.803$^{***}$ |
| | (5.949) | (5.950) | (5.844) |
| Urban | 1,119.069$^{***}$ | 1,137.706$^{***}$ | 1,178.333$^{***}$ |
| | (74.443) | (72.873) | (71.542) |
| Female | -469.414$^{***}$ | 204.542$^{***}$ | 231.176$^{***}$ |
| | (53.010) | (53.892) | (52.906) |
| Age | 8.497$^{***}$ | 44.889$^{***}$ | 37.242$^{***}$ |
| | (1.994) | (2.104) | (2.073) |
| Oe | 256.192$^{***}$ | -2.696 | 48.420$^{***}$ |
| | (7.797) | (9.460) | (9.361) |
| dij*Eij | | | 7.314$^{***}$ |
| | | | (0.170) |
| Observations | 49,356 | 49,356 | 49,356 |
| $R^2$ | 0.188 | 0.222 | 0.250 |
| Adjusted $R^2$ | 0.186 | 0.220 | 0.248 |
| Residual Std. Error | 5,355.243 (df = 49228) | 5,242.230 (df = 49227) | 5,146.001 (df = 49226) |
| *Note:* | | | $^{*}$p<0.1; $^{**}$p<0.05; $^{***}$p<0.01 |

*Figure 3: Regression of Economic distance on social distance across individuals*
*Note: Fixed effects are controlled for similar to previous regression*

Above results indicate that social distance is indeed significantly associated with economic distance between occupations. However, the association seems to be positive only when it is accompanied by a simultaneous increase in the education of an individual. An interaction between $d_{ij}$ and social groups yield similar results[7], except that given an education level for an individual, OBCs and Adivasis show a significant negative effect of social distance on economic distance. Predominant dependence on agriculture for majority of individuals within these groups may be one reason that could explain such a negative effect. A movement out of agriculture even if it's a greater social distance covered

[7] Results not shown due to paucity of space.

(since agriculture related occupations are on average located farther away from all other occupations), nevertheless may cover lesser economic distance.

## Conclusion

Overall above results highlight the importance of education, in social mobility, and also in translating the social movements out of traditional occupations to better economic prospects. There seems to be variations in the extent of mobility between occupations, across different social groups. In addition, the relationship between social distance and economic distance also seems to be greatly offset by the social group factor. These results necessitate analysis of stratification patterns separately for each social group. The social distances between occupations, community structures formed, and the differences between such structures along education levels, economic capital, and gender segregation, may be entirely different across social groups. Overall, factors like education, geography, gender, age, social group, household assets, and initial conditions of the household, seem to determine the prospective stability or flux within the broader social stratification patterns in the country. While the above framework of analysis may not be accurate or fool proof, it helps us to inquire into the dimensions along which one can possibly observe the broad social stratification patterns and consequently the structure of social inequality.

## Appendix:

Occupational network construction (in Detail):

Consider the following household information of 4 households an example, found in typical cross-sectional surveys.

| HH | PERSON ID | Occupation | Education | Age | HH | PERSON ID | Occupation | Education | Age |
|---|---|---|---|---|---|---|---|---|---|
| 1 | 1 | "61" | 1 | 22 | 3 | 1 | "35" | 5 | 25 |
| | 2 | "95" | 2 | 35 | | 2 | "22" | 6 | 35 |
| | 3 | "23" | 6 | 23 | 4 | 1 | "23" | 6 | 19 |
| 2 | 1 | "42" | 5 | 56 | | 2 | "42" | 5 | 21 |
| | 2 | "35" | 6 | 64 | | 3 | "35" | 6 | 34 |
| | 3 | "61" | 1 | 22 | | 4 | "61" | 1 | 52 |
| | 4 | "95" | 3 | 19 | | 5 | "95" | 3 | 22 |

Information from the sample survey, can be represented as a valued incidence matrix, incorporating household weights as follows (represented by say, $A$).

| HH \ Occ | "61" | "95" | "23" | "42" | "35" | "22" | HH Weight $(W_k)$ |
|---|---|---|---|---|---|---|---|
| 1 | 1 | 1 | 1 | 0 | 0 | 0 | 100 |
| 2 | 1 | 1 | 0 | 1 | 1 | 0 | 200 |
| 3 | 0 | 0 | 0 | 0 | 1 | 1 | 300 |
| 4 | 1 | 1 | 1 | 1 | 1 | 1 | 400 |

Following Borgatti et.al, and accounting for household weights, we can project the bimodal incidence matrix to one mode matrix (say $U_{\{c \times c\}}$) as follows. If total households are indexed by $h$ and occupations are indexed by $c$, then in an occupation by occupation one modal matrix $(c \times c)$, we would have $ij^{th}$element given by,

$$U_{ij} = \sum_{k=1}^{h} A_{ki} A_{kj} W_k$$

Each element in the one mode matrix $U$ therefore, represents the number of households (weighted) with members present in both corresponding row and column occupations. Diagonal elements: $U_{kk}$ represent the total number of households with at least one member, present in the occupation corresponding to index $k$. If total number of households with at least one member associated with occupation $l$ (treat this as row/column index in $U$) are $h_l$ $(= U_{ll})$ and those with at least one member in occupation $m$ are $h_m$ $(= U_{mm})$, then the expected number of households that can have members in both $l$ and $m$, under the assumption of independence shall be $\frac{h_l \times h_m}{h}$. We now perform the following transformation on $U$ to $V$, such that,

$$V_{ij} = \left(U_{ij} - \frac{h_i \times h_j}{h}\right) \Big/ \left(\frac{h_i \times h_j}{h}\right)$$

$V_{ij}$ represents the strength of connection between occupation $i$ and occupation $j$, or in other words, it can be treated as the inverse of social distance between the two occupations. This method essentially subtracts expected volume of social connections between occupations when realized through chance from the observed volume of social connections between two occupations, and further divides it with the expected value at chance. Such normalization is the basic component of all chi-square indices.

NCO-1968 Classification of occupations

NATIONAL CLASSIFICATION OF OCCUPATIONS-1968
DIVISION 0-1: PROFESSIONAL, TECHNICAL AND RELATED WORKERS
Group 00 Physical Scientists
01 Physical Science Technicians
02 Architects, Engineers, Technologists and Surveyors
03 Engineering Technicians
04 Aircraft and Ships Officers
05 Life Scientists
06 Life Science Technicians
07 Physicians and Surgeons (Allopathic Dental and Veterinary Surgeons)
08 Nursing and other Medical and Health Technicians
09 Scientific, Medical and Technical Persons, Other
10 Mathematicians, Statisticians and Related Workers
11 Economists and Related Workers
12 Accountants, Auditors and Related Workers
13 Social Scientists and Related Workers
14 Jurists
15 Teachers
16 Poets, Authors, Journalists and Related Workers
17 Sculptors, Painters, Photographers and Related Creative Artists
18 Composers and Performing Artists
19 Professional Workers, n.e.c.
DIVISION 2: ADMINISTRATIVE, EXECUTIVE AND MANAGERIAL WORKERS
Group 20 Elected and Legislative Officials
21 Administrative and Executive Officials Government and Local Bodies

22 Working Proprietors, Directors and Managers, Wholesale and Retail Trade
23 Directors and Managers, Financial Institutions
24 Working Proprietors, Directors and Managers Mining, Construction, Manufacturing and Related Concerns
25 Working Proprietors, Directors, Managers and Related Executives, Transport, Storage and Communication
26 Working Proprietors, Directors and Managers, Other Service
29 Administrative, Executive and Managerial Workers, n.e.c.

DIVISION3: CLERICAL AND RELATED WORKERS

Group 30 Clerical and Other Supervisors
31 Village Officials
32 Stenographers, Typists and Card and Tape Punching Operators
33 Book-keepers, Cashiers and Related Workers
34 Computing Machine Operators
35 Clerical and Related Workers, n.e.c.
36 Transport and Communication Supervisors
37 Transport Conductors and Guards
38 Mail Distributors and Related Workers
39 Telephone and Telegraph Operators

DIVISION 4: SALES WORKERS

Group 40 Merchants and Shopkeepers, Wholesale and Retail Trade
41 Manufacturers, Agents
42 Technical Salesmen and Commercial Travellers
43 Salesmen, Shop Assistants and Related Workers
44 Insurance, Real Estate, Securities and Business Service Salesmen and Auctioneers
45 Money Lenders and Pawn Brokers
49 Sales Workers, n.e.c.

DIVISION 5: SERVICE WORKERS

Group 50 Hotel and Restaurant Keepers
51 House Keepers, Matron and Stewards (Domestic and Institutional)
52 Cooks, Waiters, Bartenders and Related Worker (Domestic and Institutional)
53 Maids and Other House Keeping Service Workers n.e.c.
54 Building Caretakers, Sweepers, Cleaners and Related Workers
55 Launderers, Dry-cleaners and Pressers
56 Hair Dressers, Barbers, Beauticians and Related Workers
57 Protective Service Workers
59 Service Workers, n.e.c.

DIVISION 6: FARMERS, FISHERMEN, HUNTERS, LOGGERS AND RELATED WORKERS

Group 60 Farm Plantation, Dairy and Other Managers and Supervisors
61 Cultivators
62 Farmers other than Cultivators
63 Agricultural Labourers
64 Plantation Labourers and Related Workers
65 Other Farm Workers
66 Forestry Workers
67 Hunters and Related Workers
68 Fishermen and Related Workers

DIVISION 7-8-9: PRODUCTION AND RELATED WORKERS, TRANSPORT EQUIPMENT OPERATORS AND LABOURERS

Group 71 Miners, Quarrymen, Well Drillers and Related Workers
72 Metal Processors
73 Wood Preparation Workers and Paper Makers
74 Chemical Processors and Related Workers

75 Spinners, Weavers, Knitters, Dyers and Related Workers
76 Tanners, Fellmongers and Pelt Dressers
77 Food and Beverage Processors
78 Tobacco Preparers and Tobacco Product Makers
79 Tailors, Dress Makers, Sewers, Upholsterers and Related Workers
80 Shoe makers and Leather Goods Makers
81 Carpenters, Cabinet and Related Wood Workers
82 Stone Cutters and Carvers
83 Blacksmiths, Tool Makers and Machine Tool Operators
84 Machinery Fitters, Machine Assemblers and Precision Instrument Makers (except Electrical)
85 Electrical Fitters and Related Electrical and Electronic Workers
86 Broadcasting Station and Sound Equipment Operators and Cinema Projectionists
87 Plumbers, Welders, Sheet Metal and Structural Metal Preparers and Erectors
88 Jewellery and Precious Metal Workers and Metal Engravers (Except Printing)
89 Glass Formers, Potters and Related Workers
90 Rubber and Plastic Product Makers
91 Paper and Paper Board Products Makers
92 Printing and Related Workers
93 Painters
94 Production and Related Workers, n.e.c.
95 Bricklayers and Other Constructions Workers
96 Stationery Engines and Related Equipment Operators, Oilers and Greasers
97 Material Handling and Related Equipment Operators, Loaders and Unloaders
98 Transport Equipment Operators
99 Labourers, n.e.c.

DIVISION X WORKERS NOT CLASSIFIED BY OCCUPATION